\documentstyle[eqsecnum,aps]{revtex}
\begin{document}
\draft
\title{Solution to the King's problem with observables being not mutually complementary}
\author{Minoru HORIBE, Akihisa HAYASHI and Takaaki HASHIMOTO}
\address{
Department of Applied Physics\\
Fukui University, Fukui 910, Japan
}
\date{\today}
\maketitle
%-----------------<<abstract>>--------------------------------------------------------
\begin{abstract}
We investigate the King's problem of the measurement of operators
$\vec{n}_k \nobreak \cdot \nobreak \vec{\sigma}\;(k=1,2,3)$ instead of the three
Cartesian components $\sigma_x$, $\sigma_y$ and $\sigma_z$ of the spin operator $\vec{\sigma}$.
Here, $\vec{n}_k$ are three-dimensional real unit vectors.
We show the condition over three vectors $\vec{n}_k$
to ascertain the result for measurement of any one of these operators.
\end{abstract}
%-----------------<<abstract>>--------------------------------------------------------
\section{Introduction}
In the context of giving the method for inferring the outcome for measurement of
any one of three Cartesian components of spin with certainty,
Lev Vaidman, Yakir Aharonov and David Albert\cite{AAV87} introduced
the problem which is known later as the King's problem of spin-$\frac{1}{2}$ particle;  
\begin{description}
\item[Step 1]Alice sends a spin-$\frac{1}{2}$ particle to Bob.
\item[Step 2]Bob chooses any observable of $\sigma_x$, $\sigma_y$ and $\sigma_z$, and measures it
for the particle received to obtain the result $\beta(=\pm1)$. After that, Bob sends the particle back to Alice.
\item[Step 3] Alice carries out some measurements for the particle,
before Bob tells her which observable was chosen. 
From this information and the result of the measurement, Alice infers the value $\beta$ with certainty.
\end{description}
The solution using entangled state of two particles was given in that paper.

Three operators $\sigma_x$, $\sigma_y$ and $\sigma_z$ are complete in the sense that
the density matrix under consideration is uniquely determined from the probabilities for
finding the eigenstates of these operators. In addition to completeness, these operators are
mutually complementary, namely,
the eigenstates of each operator of them form mutually unbiased bases (MUB),
\[
 |\langle \beta, \vec{e}_x| \beta', \vec{e}_y \rangle|
=|\langle \beta, \vec{e}_y| \beta', \vec{e}_z \rangle|
=|\langle \beta, \vec{e}_z| \beta', \vec{e}_x \rangle|=\frac{1}{\sqrt{2}},
\]
where
$| \beta, \vec{e}_x \rangle$, $| \beta, \vec{e}_y \rangle$ and $| \beta, \vec{e}_z \rangle$
are eigenstates with eigenvalue $\beta(=\pm1)$ for observables
$\sigma_x$, $\sigma_y$ and $\sigma_z$, respectively. 

When we try to extend this to the problem in $D$ dimensional Hilbert space,
at least $D+1$ noncommuting observables are required for complete state determinations, so that, we need $D+1$
mutually unbiased bases. 
However only when the dimension of the Hilbert space is prime power,
$D+1$ mutually unbiased bases are obtained
\cite{WooF}, and we have some evidence that the number of MUB is less than $D+1$ for the case where $D$ is not equal to prime power\cite{ZAU}\cite{KLARO}\cite{ARCH}\cite{WOO04}.
For this reason, the solutions for King's problem in prime power dimensional Hilbert spaces\cite{ABE01}\cite{EA02}\cite{Ara03} are found. 

Ben-Menahem\cite{BenM89} investigated more general case for spin-$\frac{1}{2}$ particle where three observables $\vec{n}_k\cdot\vec{\sigma}\;\;(k=1,2,3)$ are used in the step 2, instead of $\sigma_x$, $\sigma_y$ and $\sigma_z$,
and Alice makes projective measurement in the step 3.  Here $\vec{n}_k$ is a real unit vector
and linearly independent of but not orthogonal to each other. These operators
are complete, but a collection of orthonormal bases formed by the eigenstates of these operators is not MUB.

In this paper, we consider the same case as he did except that the POVM measurement is made at step 3. Our method is simpler than his, although we obtain the same results.  Comparison will be made in Sec III.

\section{modified king's problem}
We try to find the solution for the modified king's problem which is obtained by exchanging three observables $\sigma_x$, $\sigma_y$ and $\sigma_z$
in the original king's problem for $\vec{n}_{k}\cdot\vec{\sigma}\;(k=1,2,3)$, following the procedure in introduction.

\noindent{\bf step 1}

Alice prepares the entangled state $|\Psi_0\rangle$ of two particles with spin $\frac{1}{2}$
\[
|\Psi_0\rangle =\frac{1}{\sqrt{2}}
\left( |+1, \vec{e}_z \rangle \otimes |-1, \vec{e}_z \rangle
      -|-1, \vec{e}_z \rangle \otimes |+1, \vec{e}_z  \rangle \right),
\]
where $| \pm 1, \vec{e}_z \rangle$ is eigenstate of the operator $\sigma_z$ with eigenvalues $\pm1$.
Since this state is a singlet state and invariant under the rotation, we can represent $|\Psi_0\rangle$ in the same form
by using the eigenstates $|\pm 1, \vec{n}_k \rangle$ with eigenvalues $\pm1$ of the operator $\vec{n}_k\cdot\vec{\sigma}$;
\[
|\Psi_0\rangle=\frac{1}{\sqrt{2}}
\left( |+1, \vec{n}_k \rangle \otimes |-1, \vec{n}_k \rangle
      -|-1, \vec{n}_k \rangle \otimes |+1, \vec{n}_k  \rangle \right).
\]
Alice sends the second particle to Bob.

\noindent{\bf step 2}

Bob chooses any one of three observables $\vec{n}_k\cdot\vec{\sigma}\;(k=1,2,3)$. Bob gets the value
$\beta$ from the measurements of it and sends this particle back to Alice. Then, Alice has two particles which are in the state,
\begin{equation}
|-\beta, \vec{n}_k \rangle \otimes |\beta, \vec{n}_k \rangle=\frac{1}{\sqrt{2}}
\left(|\vec{n}_k\rangle-\beta|\Psi_0\rangle \right),
\label{Bobst}
\end{equation}
where $|\vec{n}_k\rangle$ is given by a linear combination whose coefficients are equal to
components of the vector $\vec{n}_k$,
\[
|\vec{n}_k\rangle=(n_k)_x|X\rangle+(n_k)_y|Y\rangle+(n_k)_z|Z\rangle,
\]
and $|X\rangle$, $|Y\rangle$ and $|Z\rangle$ are defined by
\begin{eqnarray*}
|X\rangle&=&\frac{i}{\sqrt{2}}\left(|+1,\vec{e}_z\rangle \otimes |+1,\vec{e}_z\rangle
                                   +|-1,\vec{e}_z\rangle \otimes |-1,\vec{e}_z\rangle\right),\\
|Y\rangle&=&\frac{1}{\sqrt{2}}\left(|+1,\vec{e}_z\rangle \otimes |+1,\vec{e}_z\rangle
                                   -|-1,\vec{e}_z\rangle \otimes |-1,\vec{e}_z\rangle\right),\\
|Z\rangle&=&\frac{1}{\sqrt{2}}\left(|+1,\vec{e}_z\rangle \otimes |-1,\vec{e}_z\rangle
                                   +|-1,\vec{e}_z\rangle \otimes |+1,\vec{e}_z\rangle\right).
\end{eqnarray*}

\noindent{\bf step 3}

In this step, it is assumed that Alice makes a POVM measurement, because a POVM measurement  
is more general than a projective measurement. 
We can adopt the same 
strategy as the original king's problem if there is a POVM set such that, for each $k\;(k=1,2,3)$, the expectation value of an element of the POVM set in the one of two states
$|-\beta, \vec{n}_k \rangle \otimes |\beta, \vec{n}_k \rangle \; (\beta=\pm1)$ is equal to zero
and that the expectation values of the same element in the other is not equal to zero.
Since the most general POVM set like this needs $8=2^3$ elements, as is shown in the table \ref{exp1}, we consider the king's problem in which Alice's measurement is described by
the POVM set $\{E_{K}\;\;(K=A,B,\cdots, H)\}$
\begin{eqnarray}
&& \sum_{K=A}^{H}E_K={\bf 1}_{4}, \label{cndPOVM1}\\
&& E_K \geq 0\;\;(K=A,B,\cdots,H),\label{cndPOVM2}
\end{eqnarray}
where ${\bf 1}_{4}$ is identity operator on the Hilbert space of the two particles under consideration and the expectation values of the elements $E_K$ for the state in Alice's hand
are given in the table \ref{exp1}.  
It is clear that Alice can infer the value $\beta$ with certainty.
For example, when Alice gets the outcome related to POVM element $E_A$ and is told that Bob
chose the observable $\vec{n}_1\cdot\vec{\sigma}$, Alice says that $\beta=-1$ since the
probability for the outcome related to POVM element $E_A$ of a measurement performed on the
state $|-1, \vec{n}_1 \rangle \otimes |+1, \vec{n}_1 \rangle$ corresponding to $\beta=+1$ is zero. Similarly,
for other cases, Alice can infer correct $\beta$.

\begin{table}
 \caption{the probability  for the outcome related to $K (K=A,\cdots,H)$}
\begin{center}
\begin{tabular}{cc||cccc|cccc}
    Bob's choice & $\beta$ & $A$ & $B$ & $C$ & $D$ & $E$ & $F$ & $G$ & $H$ \\
\hline
\hline
    $\vec{n}_1\cdot\vec{\sigma}$ &$+1$& $0$ & $0$ &nonzero&nonzero&nonzero&nonzero& $0$ & $0$ \\
    $\vec{n}_1\cdot\vec{\sigma}$ &$-1$&nonzero&nonzero& $0$ & $0$ & $0$ & $0$ &nonzero&nonzero\\
\hline
   $\vec{n}_2\cdot\vec{\sigma}$ &$+1$ & $0$ &nonzero& $0$ &nonzero&nonzero& $0$ &nonzero& $0$ \\
   $\vec{n}_2\cdot\vec{\sigma}$ &$-1$ &nonzero& $0$ &nonzero& $0$ & $0$ &nonzero& $0$ &nonzero\\
\hline
   $\vec{n}_3\cdot\vec{\sigma}$ &$+1$ & $0$ &nonzero&nonzero& $0$ &nonzero& $0$ & $0$ &nonzero\\
   $\vec{n}_3\cdot\vec{\sigma}$ &$-1$ &nonzero& $0$ & $0$ &nonzero& $0$ &nonzero&nonzero& $0$ \\
\end{tabular}
\end{center}
\label{exp1}
\end{table}

Now we find the POVM set $\{E_{K}\;\;(K=A,B,\cdots, H)\}$. First, we consider the operator $E_{A}$.
As the operator $E_{A}$ is positive, we can have the operator $a_A$ such that
\[
E_A=a_A^{\dagger}a_A.
\]
From the table \ref{exp1}, this operator $a_A$ should satisfy three conditions,
\begin{eqnarray*}
&&a_A(|-1, \vec{n}_1 \rangle \otimes |+1, \vec{n}_1 \rangle)
=\frac{1}{\sqrt{2}}a_A
\left(|\vec{n}_1\rangle-(+1)|\Psi_0\rangle \right)=0,\\
&&a_A(|-1, \vec{n}_2 \rangle \otimes |+1, \vec{n}_2 \rangle)
=\frac{1}{\sqrt{2}}a_A
\left(|\vec{n}_2\rangle-(+1)|\Psi_0\rangle \right)=0,\\
&&a_A(|-1, \vec{n}_3 \rangle \otimes |+1, \vec{n}_3 \rangle)
=\frac{1}{\sqrt{2}}a_A
\left(|\vec{n}_3\rangle-(+1)|\Psi_0\rangle \right)=0,
\end{eqnarray*}
where we used the eq.(\ref{Bobst}).
 As three states $|\vec{n}_k\rangle-(+1)|\Psi_0\rangle\;\;(k=1,2,3)$ are
linearly independent of each other in $4$-dimensional Hilbert space
owing to linear independence of three vectors $\vec{n}_k$, 
using state $\langle \Psi_0|+\sum_{k=1}^{3}(S^{(A)}M^{-1})_{k}\langle \vec{n}_k|$ orthogonal to these three states,
the operator $a_A$ is written in the form,
\begin{equation}
a_A=|\Phi_A \rangle (\langle \Psi_0|+\sum_{k=1}^{3}(S^{(A)}M^{-1})_{k}\langle \vec{n}_k|), 
\label{aAop}
\end{equation}
where $|\Phi_A \rangle $ is an undetermined state from these conditions, $S^{(A)}$ is a three dimensional real vector
\[
S^{(A)}=(+1,+1,+1),
\]
and $M$ is a $3\times3$ matrix whose $(i,j)$-component is given by inner product between $\vec{n}_i$ and $\vec{n}_j$
and invertible because three vectors $\vec{n}_i$ are lineally independent of each other.
Thus we get POVM element $E_A$
\[
E_A=a_A^{\dagger}a_A=C_A(|\Psi_0 \rangle+\sum_{k=1}^{3}(S^{(A)}M^{-1})_{k}|\vec{n}_k\rangle)
                        (\langle \Psi_0|+\sum_{k=1}^{3}(S^{(A)}M^{-1})_{k}\langle \vec{n}_k|),
\]
where $C_A$ is a nonnegative constant
\[
C_A=\frac{1}{2}\langle \Phi_A|\Phi_A \rangle.
\]
Similarly, $E_K$ is restricted to the form
\begin{equation}
E_K=C_K(|\Psi_0 \rangle+\sum_{k.l=1}^{3}(S^{(K)}M^{-1})_{k}|\vec{n}_k\rangle)
                        (\langle \Psi_0|+\sum_{k.l=1}^{3}(S^{(K)}M^{-1})_{k}\langle \vec{n}_k|)
\;\;(K=A,B,C,\cdots, H),
\label{povnek}
\end{equation}
where three dimensional vectors $(S^{(K)})_k$ are given by
\begin{eqnarray*}
&&S^{(B)}=(+1,-1,-1),\;\;\;
  S^{(C)}=(-1,+1,-1),\\
&&S^{(D)}=(-1,-1,+1),\;\;\;
  S^{(E)}=(-1,-1,-1),\\
&&S^{(F)}=(-1,+1,+1),\;\;\;
  S^{(G)}=(+1,-1,+1),\\
&&S^{(H)}=(+1,+1,-1).
\end{eqnarray*}
From the condition (\ref{cndPOVM1}) of POVM set, the constants $C_K$ satisfy equations
\begin{eqnarray}
&& \sum_{K=A}^{H}C_K=1,\label{conC1}\\
&& \sum_{K=A}^{H}C_K(S^{(K)})_l=0,\label{conC2}\\
&& \sum_{K=A}^{H}C_K(S^{(K)})_k(S^{(K)})_l=(M)_{kl}.\label{conC3}
\end{eqnarray}
In order to get these equations, we used the expansion of identity matrix in
a set $\{|[\Psi_0\rangle,\;\;|\vec{n}_k\rangle\;\;(k=1,2,3)\}$
\[
{\bf 1}_4=|\Psi_0\rangle \langle \Psi_0|+\sum_{k,l}^3|\vec{n}_k\rangle (M^{-1})_{kl}\langle \vec{n}_l|.
\]
From these equations (\ref{conC1})$\sim$(\ref{conC3}), we have seven independent equations for eight variables $C_K$;
\begin{eqnarray}
(C_A + C_E) + (C_B + C_F) + (C_C + C_G) + (C_D + C_H) &=& 1, \nonumber\\
(C_A + C_E) - (C_B + C_F) - (C_C + C_G) + (C_D + C_H) &=& \vec{n}_1\cdot\vec{n}_2,  \nonumber\\
(C_A + C_E) - (C_B + C_F) + (C_C + C_G) - (C_D + C_H) &=& \vec{n}_1\cdot\vec{n}_3,\nonumber\\
(C_A + C_E) + (C_B + C_F) - (C_C + C_G) - (C_D + C_H) &=& \vec{n}_2\cdot\vec{n}_3,\label{conCK}\\
(C_A - C_E) + (C_B - C_F) - (C_C - C_G) - (C_D - C_H) &=& 0,\nonumber\\
(C_A - C_E) - (C_B - C_F) + (C_C - C_G) - (C_D - C_H) &=& 0,\nonumber\\
(C_A - C_E) - (C_B - C_F) - (C_C - C_G) + (C_D - C_H) &=& 0,\nonumber
\end{eqnarray}
and we get a solution with one parameter $r$,
\begin{eqnarray*}
C_A&=&\frac{1}{8}(1+r +\vec{n}_1\cdot\vec{n}_2
                      +\vec{n}_1\cdot\vec{n}_3
                      +\vec{n}_2\cdot\vec{n}_3),\\
C_B&=&\frac{1}{8}(1+r -\vec{n}_1\cdot\vec{n}_2
                      -\vec{n}_1\cdot\vec{n}_3
                      +\vec{n}_2\cdot\vec{n}_3),\\
C_C&=&\frac{1}{8}(1+r -\vec{n}_1\cdot\vec{n}_2
                      +\vec{n}_1\cdot\vec{n}_3
                      -\vec{n}_2\cdot\vec{n}_3),\\
C_D&=&\frac{1}{8}(1+r +\vec{n}_1\cdot\vec{n}_2
                      -\vec{n}_1\cdot\vec{n}_3
                      -\vec{n}_2\cdot\vec{n}_3),\\
C_E&=&\frac{1}{8}(1-r +\vec{n}_1\cdot\vec{n}_2
                      +\vec{n}_1\cdot\vec{n}_3
                      +\vec{n}_2\cdot\vec{n}_3),\\
C_F&=&\frac{1}{8}(1-r -\vec{n}_1\cdot\vec{n}_2
                      -\vec{n}_1\cdot\vec{n}_3
                      +\vec{n}_2\cdot\vec{n}_3),\\
C_G&=&\frac{1}{8}(1-r -\vec{n}_1\cdot\vec{n}_2
                      +\vec{n}_1\cdot\vec{n}_3
                      -\vec{n}_2\cdot\vec{n}_3),\\
C_H&=&\frac{1}{8}(1-r +\vec{n}_1\cdot\vec{n}_2
                      -\vec{n}_1\cdot\vec{n}_3
                      -\vec{n}_2\cdot\vec{n}_3).
\end{eqnarray*}
Unfortunately, since the coefficients $C_K$ is nonnegative, all three unit vectors $\vec{n}_k$
which are linearly independent of each other are not permitted.  However, we can easily see that we get this POVM set if linearly independent unit vectors $\vec{n}_k,(k=1,2,3)$ satisfy the following inequality
\begin{equation}
1>|\vec{n}_1\cdot\vec{n}_2|+|\vec{n}_2\cdot\vec{n}_3|+|\vec{n}_3\cdot\vec{n}_1|,
\label{fcon1}
\end{equation}
and it is clear that there are three vectors $\vec{n}_k$ satisfying this inequality. When we
express the solution $C_K$ with different forms  
\begin{eqnarray*}
C_A&=& \frac{1}{16}\left\{| \vec{n}_1+\vec{n}_2+\vec{n}_3|^2+2r-1\right\},\\
C_B&=& \frac{1}{16}\left\{|-\vec{n}_1+\vec{n}_2+\vec{n}_3|^2+2r-1\right\},\\
C_C&=& \frac{1}{16}\left\{| \vec{n}_1-\vec{n}_2+\vec{n}_3|^2+2r-1\right\},\\
C_D&=& \frac{1}{16}\left\{| \vec{n}_1-\vec{n}_2-\vec{n}_3|^2+2r-1\right\},\\
C_E&=& \frac{1}{16}\left\{| \vec{n}_1+\vec{n}_2+\vec{n}_3|^2-2r-1\right\},\\
C_F&=& \frac{1}{16}\left\{|-\vec{n}_1+\vec{n}_2+\vec{n}_3|^2-2r-1\right\},\\
C_G&=& \frac{1}{16}\left\{| \vec{n}_1-\vec{n}_2+\vec{n}_3|^2-2r-1\right\},\\
C_H&=& \frac{1}{16}\left\{| \vec{n}_1-\vec{n}_2-\vec{n}_3|^2-2r-1\right\},
\end{eqnarray*}
we can find the necessary and sufficient condition 
which guarantees that these variables $C_K$ are nonnegative;
\begin{equation}
|\vec{n}_1 \pm \vec{n}_2 \pm \vec{n}_3| \ge 1,
\label{fcon2}
\end{equation}
for all combinations of signs in front of the second and the third terms of the left hand side.

\section{Discussion and Summary}
In this paper we considered the modified king's problem that Bob chooses any one from three observables $\vec{n}_k\cdot\vec{\sigma}$, instead of $\sigma_x$, $\sigma_y$ and $\sigma_z$
and that he makes measurement of it in step 2.  We showed that, if linearly independent unit
vectors $\vec{n}_k$ satisfy the inequality (\ref{fcon2}), Alice can infer the result of Bob's
measurement from the outcome for the measurement of POVM $E_K(K=A,B,\cdots,H)$ and
the information of Bob's choice with certainty. 

Ben-Menahem\cite{BenM89} considered the same model with projective measurement in step 3 and concluded that
no ineqalities were imposed on $\vec{n}_l\cdot\vec{n}_k$ beyond the geometric ones.
He derived the equations for the coefficients $b_A$ of expansion for initially prepared state in eigenstates
of the observable Alice measures in the final step;
\begin{eqnarray*}
&&\sum_{A}d_A=1,\\
&&\sum_{A}\epsilon^{(l)}_A\epsilon^{(k)}_Ad_A=\vec{n}_l\cdot\vec{n}_k,
\end{eqnarray*}
where $d_A=|b_A|^2$ and $\epsilon^{(l)}_A$, which is a factor related to Alice's strategy, takes $\pm1$.
The first equation is normalization condition for initially prepared state.
Replacing $d_A$ and $\epsilon^{(l)}_A$ with $C_K$ and $S^{(K)}_l$, respectively, we can see that these equations become the same as equations (\ref{conC1}) and (\ref{conC3}) we
solved in Sec II, although the physical meaning of $d_A$ are different from one of $C_K$.
It is shown that the vectors $\vec{n}_{l}$
have to satisfy our inequalities (\ref{fcon2}), from these equations and positivity condition for $d_A$. For $|\sum_{k=1}^3\vec{n}_k|$ we have
\begin{eqnarray*}
\left(\sum_{k=1}^3\vec{n}_k\right)\cdot
\left(\sum_{l=1}^3\vec{n}_l\right)&=&\sum_{k,l=1}^3\sum_{A}\epsilon^{(l)}_A\epsilon^{(k)}_Ad_A,\\
&=&\sum_{A}\left(\sum_{l=1}^3\epsilon^{(l)}_A\right)
           \left(\sum_{k=1}^3\epsilon^{(k)}_A\right)d_A,\\
&\ge&\sum_{A}d_A=1.
\end{eqnarray*}
Here we used the inequalities
\[
\left|\sum_{l=1}^3\epsilon^{(l)}_A \right| \ge 1,\;\;\;d_A \ge 0.
\]
Similarly other inequalities are obtained.
Indeed, after tedious but not difficult calculation, we can see that our inequalities (\ref{fcon2}) for vectors $\vec{n}_k$ are equivalent to positivity conditions for $d_A$. If the conclusion in his paper\cite{BenM89} were right, any three
vectors $\vec{n}_k$ would satisfy our inequalities (\ref{fcon2}). However
it is not difficult for us to find configuration of three vectors $\vec{n}_l$ which do not satisfy our inequalities .
When, as these vectors $\vec{n}_k$, we choose vectors obtained by rotating three vectors on $x-y$ plane, such that
angles between each other are equal to $\frac{2\pi}{3}$, by small angle toward $z$-axis, inner products become
\[
\vec{n}_1\cdot \vec{n}_2=-\frac{1}{2}+\delta_{12},\;\;
\vec{n}_2\cdot \vec{n}_3=-\frac{1}{2}+\delta_{23},\;\; 
\vec{n}_3\cdot \vec{n}_1=-\frac{1}{2}+\delta_{31},\;\;(\delta_{12}+\delta_{23}+\delta_{31}< 1)
\]
and we get
\[
(\vec{n}_1+\vec{n}_2+\vec{n}_3)\cdot (\vec{n}_1+\vec{n}_2+\vec{n}_3)=\delta_{12}+\delta_{23}+\delta_{31}<1.
\]

Owing to completeness of the observable $\vec{n}_k\cdot\vec{\sigma}$, we considered the case
where three unit vectors $\vec{n}_k$ are linearly independent. We investigate the kings problem
defined by three vectors $\vec{n}_k$ that are linearly dependent but that are not parallel. 
Without losing the generality, it is supposed that the vectors $\vec{n}_k$ satisfy 
\begin{equation}
\vec{n}_3=x\vec{n}_1+y\vec{n}_2,\;\;(x,y \neq 0).
\label{depcon1}
\end{equation}
As these vectors are unit vector, we have
\begin{equation}
(x \pm y)^2=1-2xy(\vec{n}_1\cdot\vec{n}_2 \mp 1)\;\; {\rm or}\;\;
(x \pm y)^2\neq 1.
\label{depcon2}
\end{equation}
We return to the condition for $a_K$ defined by $E_K=a_K^{\dagger}a_K$ 
\begin{equation}
a_K(|-(S^{(K)})_k, \vec{n}_k \rangle \otimes |(S^{(K)})_k, \vec{n}_k \rangle)
=\frac{1}{\sqrt{2}}a_K
\left(|\vec{n}_k\rangle-(S^{(K)})_k |\Psi_0\rangle \right)=0.
\label{conak}
\end{equation}
Using eq. (\ref{depcon1}), we have
\[
\left\{(S^{(K)})_3-x(S^{(K)})_1-y(S^{(K)})_2\right\}a_K|\Psi_0\rangle=0
\;\;\;{\rm or}\;\;\;a_K|\Psi_0\rangle=0.
\]
Here we used the eq.(\ref{depcon2}).  The condition (\ref{conak}) is  rewritten in
the equations
\begin{eqnarray*}
&&a_K|\Psi_0\rangle=0,  \\
&&a_K|\vec{n}_1\rangle=a_K|\vec{n}_1\rangle=0.
\end{eqnarray*}
Therefore we can express all $E_K$ in
the state $|\vec{n}_1 \times \vec{n}_2 \rangle$ orthogonal to states
$|\Psi_0\rangle$, $|\vec{n}_1\rangle$ and $|\vec{n}_2\rangle$,
\[
E_K=C_K|\vec{n}_1 \times \vec{n}_2 \rangle \langle \vec{n}_1 \times \vec{n}_2|.
\]
However the set $\{E_K\}$ is not POVM because
$\displaystyle{\sum_{K=A,B,\cdots,H}E_K \neq {\bf 1}_4}$. 
Thus there is no solution to the king's problem in this case.

Can we reduce the number of elements of POVM set from 8 to 4 like the original problem?
In the equation (\ref{conCK}) of the previous section, substituting zero into the variables
$C_E$, $C_F$ ,$C_G$ and $C_H$, we have equations
\begin{eqnarray*}
C_A + C_B + C_C + C_D &=& 1,\\
C_A - C_B - C_C + C_D &=& \vec{n}_1\cdot\vec{n}_2,\\
C_A - C_B + C_C - C_D &=& \vec{n}_1\cdot\vec{n}_3,\\
C_A + C_B - C_C - C_D &=& \vec{n}_2\cdot\vec{n}_3,\\
C_A + C_B - C_C - C_D &=& 0,\\
C_A - C_B + C_C - C_D &=& 0,\\
C_A - C_B - C_C + C_D &=& 0.
\end{eqnarray*}
If and only if
these unit vectors $\vec{n}_k$ are orthogonal to each other,
the solution exists to the above equations and  different element $E_K$ of the POVM set in this solution extracts different state of an orthonormal basis in four dimensional Hilbert
space of two spin-$\frac{1}{2}$ particles.
Thus the modified king's problem results in the original one. When $C_A$, $C_B$, $C_C$ and $C_D$
are zero, we can get the same result. We can show that there is no solution to the equations
for other cases, when three vectors $\vec{n}_k$ are linearly independent,
after the tedious but not difficult calculations.

We had the solution for the king's problem using three observables which are complete but not mutually complementary for spin-$\frac{1}{2}$ particle. 
However we have not discussed the problems such that
\begin{enumerate}
\item Bob chooses any one of two observables.
\item Alice uses other entangled state.
\item The dimension of the Hilbert space is larger than two.
\end{enumerate}
We will discuss king's problem for these cases elsewhere.

%=====<< References >>==========================

\end{document}